\documentclass[aps,pra,floatfix,twocolumn,superscriptaddress,nofootinbib,showpacs]{revtex4} 

\usepackage{graphicx}
\usepackage{graphics}
\usepackage{amssymb}
\usepackage{amsmath}
\usepackage{epsfig}
\usepackage{latexsym}
\usepackage{color}

\newcommand{\ket}[1]{\left\vert{#1}\right\rangle}
\def\ketc[#1]{\vert #1 \rangle}
\def\brac[#1]{\langle #1 \vert}

\newcommand{\beq}{\begin{equation}}
\newcommand{\eeq}{\end{equation}}
\newcommand{\bqa}{\begin{eqnarray}}
\newcommand{\eqa}{\end{eqnarray}}
\newcommand{\nn}{\nonumber}

\newcommand{\erf}[1]{Eq.~(\ref{#1})}
\newcommand{\dg}{^\dagger}
\begin{document}

\title{High fidelity measurement and quantum feedback control in circuit QED}

\author{Mohan Sarovar}
\email{mohan@physics.uq.edu.au} \affiliation{Centre for Quantum
Computer Technology, and School of Physical Sciences, The
University of Queensland, St Lucia, QLD 4072, Australia}

\author{Hsi-Sheng Goan}
\email{goan@phys.ntu.edu.tw} \affiliation{Department of Physics, National Taiwan University, Taipei 106 Taiwan}

\author{T. P. Spiller}
\email{timothy.spiller@hp.com}
\affiliation{Hewlett-Packard Laboratories, Filton Road, Stoke Gifford, Bristol BS34 8QZ, United Kingdom}

\author{G. J. Milburn}
\email{milburn@physics.uq.edu.au} \affiliation{Centre for Quantum
Computer Technology, and School of Physical Sciences, The
University of Queensland, St Lucia, QLD 4072, Australia}

%------------------------------------------------------------------------------
% Abstract
%------------------------------------------------------------------------------

\begin{abstract}
Circuit QED is a promising solid-state quantum computing architecture. It also has excellent potential as a platform for quantum control -- especially quantum feedback control -- experiments. However, the current scheme for measurement in circuit QED is low efficiency and has low signal-to-noise ratio for single shot measurements. The low quality of this measurement makes the implementation of feedback difficult, and here we propose two schemes for measurement in circuit QED architectures that can significantly improve signal-to-noise, and potentially achieve quantum limited measurement. Such measurements would enable the implementation of quantum feedback protocols and we illustrate this with a simple entanglement stabilization scheme.
\end{abstract}

\maketitle

%------------------------------------------------------------------------------
% Text
%------------------------------------------------------------------------------

%Intro
%Current meas and limitations
%Proposals
%  - SET
%  - Microwave cavity
%Performance Analysis
%Application

\section{Introduction}

The emergence of quantum computing has revitalized the field of quantum control \cite{War.Rab.etal-1993, Hab.Jac.etal-2002, Wis-1994, Doh.Hab.etal-2000} in recent years. The prospects for a working quantum computer are intimately dependent on the ability to reliably control individual and coupled quantum systems. There have been several experimental proposals for architectures capable of quantum computing \cite{mikeandike} and most are in their initial stages of experimental development. During these initial stages, quantum computing experiments are either not possible or only proof of principle -- mainly due to the small number of qubits under control. However, many of these quantum computing architectures, during their initial stages of experimental development, are ideal testbeds for quantum control experiments designed to investigate the limits of quantum control, quantum chaos, and the fundamentals of quantum mechanics. 

A particularly promising group of quantum computing architectures that are well suited to quantum control experiments are the solid-state realizations of cavity quantum electrodynamics. These architectures couple superconducting electronic circuit elements, which serve as the qubits, to either electromagnetic modes \cite{Wal.Sch.etal-2004, Chi.Ber.etal-2004} or nanoelectromechanical resonators \cite{Gel.Cle-2005}, which serve as a `quantum bus' that mediates inter-qubit coupling and/or facilitates measurement of a qubit's state. Strong coupling between the qubits and the quantum bus, and the substantial amount of control afforded by these architectures over both subsystems makes these systems excellent quantum control testbeds. 

In this paper we will focus on one such implementation, that of Schoelkopf \textit{et. al.} \cite{Wal.Sch.etal-2004} at Yale University, and examine it from a quantum control perspective. This implementation couples the quantized charge degree of freedom of a Cooper pair box (CPB) \cite{Mak.Sch.etal-2001, Nak.Pas.etal-1999, Dev.Wal.etal-2004} to a harmonic oscillator mode of a microwave resonator. Both the Cooper pair box and the resonator are fabricated on an integrated circuit and thus the setup has been dubbed `circuit QED'. A number of features of this architecture make it well suited to quantum control -- especially quantum feedback control -- experiments: the fact that the CPBs (which can be viewed as artificial atoms) are localized and the high degree of electrical control over them, the strong coupling between the CPBs and the resonator which allows for a well controlled `atom'-field interaction, the fact that feedback can be applied to either the CPB or the microwave resonator mode, and of course, the integrated nature of the whole apparatus. However, the current scheme for measurement of the qubits is highly inefficient and has a poor signal to noise ratio (for single shot measurements) \cite{Wal.Sch.etal-2005}. The primary reason for this is that while the qubit and resonator are at a temperature of 20mK, the mixing electronics and local oscillator needed for amplitude measurements are at room temperature. See figure 2 in Ref. \cite{Wal.Sch.etal-2004}. The noise introduced by this room temperature mixing, and the amplification stage before it, results in a measurement signal with low single-shot signal-to-noise ratio. This low fidelity measurement scheme is a problem from a quantum control perspective because high quality measurements are essential to quantum control, especially quantum feedback control, experiments. Ideally the measurements should be quantum-limited for quantum feedback control. Motivated by this, we propose two methods for high fidelity measurement in the circuit QED architecture. The first uses an single electron transistor (SET) as a heterodyne mixer and the second performs the local oscillator mixing at the cryogenic level by the addition of a second transmission line resonator that carries a local oscillator to the circuit . 

This paper is organized as follows: sections \ref{sec:cqed} and \ref{sec:setmeas} review the circuit QED architecture of Wallraff \textit{et. al.}, and techniques for SET based measurement. Section \ref{sec:proposals} presents our two proposals for high fidelity measurements in circuit QED. Section \ref{sec:fb} describes a simple feedback protocol for deterministically creating and stabilizing entanglement between two qubits that would be made possible by such high fidelity measurements, and finally, section \ref{sec:conc} concludes with a discussion.

\section{Circuit QED}
\label{sec:cqed}
The qubits in the circuit QED architecture are split junction Cooper pair boxes \cite{Dev.Wal.etal-2004, Ast.Pas.etal-2004}. These devices can be modelled as two-level systems with the Hamiltonian \cite{Mak.Sch.etal-2001}
\begin{equation}
\label{eq:cpb_ham}
H_a = -\frac{1}{2}(E_{el}\tilde{\sigma}_z + E_J\tilde{\sigma}_x),
\end{equation}
where $E_{el}$ is the electrostatic energy and $E_J$ is the Josephson coupling energy. The Pauli matrices $\tilde{\sigma_z}$ and $\tilde{\sigma_x}$ are in the charge basis; that is, the basis states correspond to either zero or one excess Cooper-pair charges on the island. 

These CPBs can be viewed as artificial atoms with large dipole moments, and in circuit QED they are coupled to microwave frequency photons in a quasi-one dimensional transmission line cavity (a co-planar waveguide resonator) by an electric dipole interaction \cite{Wal.Sch.etal-2004}. This apparatus has a number of \textit{in situ} tunable parameters, including $E_{el}$ and $E_J$, and a choice can be made \cite{Bla.Hua.etal-2004} such that the combined Hamiltonian, for qubit and transmission line cavity, is the well-known Jaynes-Cummings Hamiltonian \cite{wallsandmilburn}:
\begin{equation}
\label{eq:cqed_ham}
H = \hbar \omega_r \big( a\dg a + \frac{1}{2} \big) + \frac{\hbar\omega_a}{2} \sigma_z + \hbar g(a\dg \sigma^- + a\sigma^+) 
\end{equation}
where $a$ ($a\dg$) is the annihilation (creation) operator for the cavity mode, $\omega_r$ is the cavity resonance frequency, $\omega_a$ is the energy splitting of the qubit, and $g$ is the coupling strength. 
Of course, the qubit energy splitting, $\omega_a$, is a function of the Cooper pair box parameters $E_{el}$ and $E_J$ \cite{Bla.Hua.etal-2004}. For typical values of the parameters in this Hamiltonian see Refs. \cite{Wal.Sch.etal-2004, Bla.Hua.etal-2004}. It should be noted here that there has been a basis change between Eqns. (\ref{eq:cpb_ham}) and (\ref{eq:cqed_ham}). In essence, we have swapped the $x$ and $z$ axes of \erf{eq:cpb_ham}, and the computational basis for our qubit has become the Josephson basis of the CPB. This choice of basis, which corresponds to operating the CPB in what is called the \textit{charge degeneracy point} ($E_{el} \sim 0$), has a number of advantages, the primary one being that the computational basis states become first-order insensitive to dephasing from offset charge noise \cite{Dev.Wal.etal-2004}. In fact, a CPB is only effective as a robust qubit at this operating point.

In the dispersive regime, where $\Delta \equiv \omega_a - \omega_r \gg 1$, the above Hamiltonian can be diagonalized to obtain \cite{Bla.Hua.etal-2004}
\begin{equation}
\label{eq:cqed_disp_ham}
H \approx \hbar \big( \omega_r + \frac{g^2}{\Delta}\sigma_z \big) a\dg a + \frac{1}{2}\hbar \big( \omega_a + \frac{g^2}{\Delta} \big) \sigma_z
\end{equation}
This form of the Hamiltonian makes evident the qubit-state dependent pull of the cavity frequency in this regime. Thus readout of the qubits in this architecture can be done by irradiating the cavity and then probing the reflected or transmitted amplitude. A quantum non-demolition (QND) measurement of the qubit in the $\sigma_z$ basis can be made by measuring the phase shift $\phi$ of a probe microwave through the cavity at frequency $\omega_r$ \cite{Bla.Hua.etal-2004}. Physically, this measurement localizes the qubit in the Josephson basis, the eigenbasis of $\sigma_z$ in \erf{eq:cqed_ham}, and the CPB remains operating at the optimal charge degeneracy point.

\begin{figure}
\includegraphics[scale=0.6]{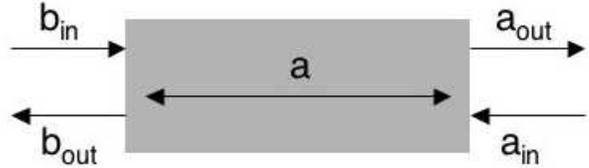}
\caption{Modeling the resonator as a two-sided leaky cavity.} \label{fig:cavity}
\end{figure}

The precise dependence of the phase of the output field on the qubit state can be seen from the input-output relations between the in and out cavity fields. If we model the resonator as a two-sided cavity with equal rates of leakage on both ends, as in figure \ref{fig:cavity}, then the relationship between the mode in the cavity and the output mode is \cite{wallsandmilburn} 
\begin{equation}
a_{out}(t) = \sqrt{\kappa}a(t) - a_{in}(t)
\label{eq:out_ic_rel}
\end{equation}
where $a_{out}$ is the output mode, $a$ is the intra-cavity mode, $a_{in}$ is the input mode at the output port, and $\kappa$ is the leakage rate. This treatment of the transmission line resonator as a cavity follows from the quantization of such quasi-one-dimensional transmission lines \cite{Lou-1990}. The mode operator $a$ of the cavity is related to the voltage and current carried by the transmission line by \cite{Lou-1990}:
\begin{eqnarray}
V(x, t) &=& \sqrt{\frac{\hbar\omega}{2Cl}}~[a(t)e^{ikx} + a(t)\dg e^{-ikx}] \nn \\
I(x,t) &=& \sqrt{\frac{C}{L}}~V(x,t)
\end{eqnarray}
where $L$ and $C$ are the inductance and capacitance per unit length of the transmission line, and $l$ is the length of the transmission line. For a given frequency, $\omega$, the propagation constant $k$ is fixed by $k=\omega\sqrt{LC}$. These relations come from solving the classical equations of motion for a lossless transmission line and quantizing the result \cite{Lou-1990}. Note that once $a(t)$ is given the solution is completely specified, and therefore we characterize the state of the transmission line completely by defining $a(t)$.

We will assume that $a_{in}(t)$, the input mode at the output port is always the vacuum because the resonator can be engineered to minimize reflection and back-scatter. Given this model, we have the following equation of motion for the intra-cavity mode \cite{wallsandmilburn}:
\begin{eqnarray}
\frac{da(t)}{dt} &=& -\frac{i}{\hbar}[a(t), H_{sys}] - \kappa a(t) + \sqrt{\kappa}a_{in}(t) + \sqrt{\kappa}b_{in}(t) \nn \\
&=& -i[a(t), (\omega_r + \frac{g^2}{\Delta} \sigma_z) a\dg a] - \kappa a(t) \nn \\ 
&& + \sqrt{\kappa}a_{in}(t) + \sqrt{\kappa}b_{in}(t)
\end{eqnarray}
where $H_{sys}$ describes the dynamics of the intra-cavity mode, and $b_{in}$ is the input mode at the input port of the resonator. Henceforth let $\chi \equiv g^2/\Delta$. If we Fourier transform this differential equation, use \erf{eq:out_ic_rel}, and rearrange, we can get a relationship between the input and output mode spectra:
\begin{eqnarray}
a_{out}(\omega) &=&  \frac{\kappa}{\kappa + i(\omega_r - \omega + \chi\sigma_z)} \big( b_{in}(\omega) + a_{in}(\omega)\big) \nn \\
&-& a_{in}(\omega)
\end{eqnarray} 
where the $\sigma_z$ in the denominator is simply a formal indication that the output spectra depends on the qubit state. So when the qubit is in a computational basis state ($\sigma_z = \pm 1$), the output mode at $\omega=\omega_r$, the cavity resonance frequency, is:
\begin{eqnarray}
a^{\pm}_{out}(\omega_r) &=&  \frac{\kappa}{\kappa \pm i\chi}b_{in}(\omega_r) \mp \frac{i\chi}{\kappa \pm i\chi}a_{in}(\omega_r)
\label{eq:inp_oup}
\end{eqnarray} 
Now, because $a_{in}(t)$ is the vacuum and the two input modes are assumed to be independednt, the second term of \erf{eq:inp_oup} will not contribute to any normally ordered moments of $a_{out}$, and we can effectively ignore it (even though it is strictly required to preserve commutation relations). Thus,
\begin{eqnarray}
\langle a^{\pm}_{out}(\omega_r)\rangle_{\mathcal{N}} = \langle \frac{\kappa^2 \mp i\kappa \chi}{\kappa^2 + \chi^2} b_{in}(\omega_r)\rangle_{\mathcal{N}} = \langle e^{i\theta^{\mp}} \tilde{b}_{in}(\omega_r) \rangle_{\mathcal{N}}
\label{eq:norm_moments}
\end{eqnarray} 
where $\langle \cdot \rangle_{\mathcal{N}}$ denotes a normally ordered moment, $\tilde{b}_{in}$ is a scaled version of $b_{in}$ and $\theta^{\mp}$ is the qubit state dependent phase difference between the input and output fields ($\tan(\theta^{\mp}) = \mp \frac{g^2}{\kappa \Delta}$), exactly the quantity we want to measure in order to perform a QND measurement of the qubit. It can be measured using the standard heterodyne technique of mixing with a local oscillator of well defined phase to an intermediate frequency. However, in current experiments this mixing is done at room temperature after amplification of the source signal, and this amplification and room temperature mixing both introduce too much noise for a high fidelity single shot measurement of the qubit state to be possible \cite{Wal.Sch.etal-2005}. It is precisely this problem that we address in this paper. In section \ref{sec:proposals} we propose two schemes for performing this QND measurement with high fidelity.

\section{Measurements using an SET}
\label{sec:setmeas}
A single electron transistor (SET) is a device that is the combination of a small capacitance metallic island and high resistance source and drain leads. Electrons can tunnel from the leads onto the island and off again. Typically, one or more `gates' are also capacitively coupled to the island, and voltages applied to these gates serve to change the electrostatic energy of the island and thus the electron tunneling rate. The typical SET operating regime is the so-called Coulomb blockade regime (which is set up by a choice of the source-drain voltage), where the source-drain current is formed by single electrons tunneling between the contacts through the island which will only accommodate one excess electron at a time. In this regime, the voltage applied to the gate(s) has a dramatic effect on the tunneling rate and thus the device becomes a highly sensitive electrometer. 

\begin{figure}
\includegraphics[scale=0.6]{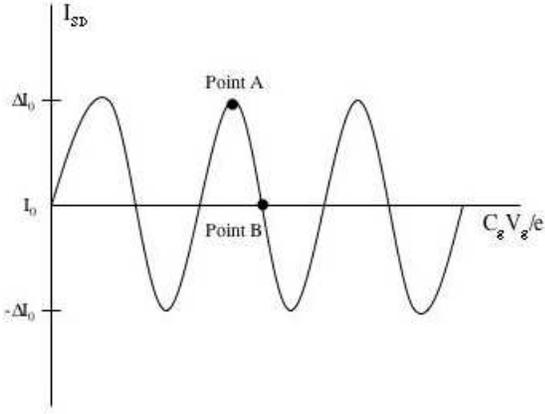}
\caption{Approximate gate voltage versus source-drain current curve for an SET operating in the strong tunneling regime. The operating point of the SET is changed by applying a DC bias voltage to the gate. The two most useful operating points for measurement are denoted by Point A and Point B on the diagram. The x-axis is dimensionless and the y-axis is in arbitrary current units.} \label{fig:set_resp}
\end{figure}

In addition to the Coulomb-blockade regime, we can also operate an SET in the strong tunneling regime, where the source-drain voltage is larger than in the Coulomb-blockade regime, leading to larger tunnel conductances. In this regime the source-drain current is less sensitive to the gate voltage; in fact for near-optimally biased source-drain voltage, the source-drain current versus gate voltage curve in this regime is well approximated by a sinusoid \cite{Kno.Yun.etal-2002, Swe.Sch.etal-2005}. This is illustrated in Figure \ref{fig:set_resp}. We will see that this strong tunneling regime is useful for performing measurements using an SET because of the well defined non-linear current-voltage behaviour. While operating in this regime, a DC bias voltage can be applied to the gates to move the operating point along the voltage-current curve. The two most useful points along this curve from a measurement perspective are also illustrated in Figure \ref{fig:set_resp}.

\subsection{The SET as a mixer}
The non-linear relationship between the gate voltage and the source-drain current of an SET makes it possible to operate it as a frequency mixer, as demonstrated by Refs. \cite{Kno.Yun.etal-2002, Swe.Sch.etal-2005, Rei.Bue-2005}. We shall use this feature of the SET extensively, and therefore in the following we will derive the exact relationship between the output current and gate voltage of an SET. 
 
Consider an SET operating in the strong tunneling regime with a voltage $V_g(t) = V_g^{DC} + V_1(t) + V_2(t)$ applied to the gate(s). We will analyze this situation because it will be the relevant scenario when we consider making measurements in circuit QED using an SET (in that case $V_1$ will correspond to the signal voltage from the cavity and $V_2$ will be a local oscillator signal). $V_G^{DC}$ is a DC bias voltage that controls the SET operating point on the current-voltage curve of Fig. \ref{fig:set_resp}. We wish to obtain an expression for the source-drain current when the SET is operating at point A on the current-voltage curve. Because of the sinusoidal dependence on voltage, we can write the source-drain current as:
\begin{equation}
I(t) = I_0 + \Delta I_0 \cos \big( \frac{2\pi C_g}{e} (V_g^{DC} + V_1(t) + V_2(t)) \big)
\end{equation}
where $V_g^{DC}$ is the DC bias voltage necessary to operate at point A. Now we will assume that $V_1(t)$ and $V_2(t)$ each have a well-defined single frequency, and expand each out in a quadrature representation:
\begin{eqnarray}
I(t) &=& I_0 + \Delta I_0 \cos \big[ \frac{2\pi C_g}{e} (V_g^{DC} + \nn \\
 && X_1(t)\cos(\Omega_1t) + Y_1(t)\sin(\Omega_1 t) + \nn \\
 && X_2(t)\cos(\Omega_2 t + \phi) + Y_2(t)\sin(\Omega_2 t + \phi)) \big]
\end{eqnarray}
where $\phi$ is the phase difference between $V_1$ and $V_2$. Henceforth, let $c_g \equiv 2\pi C_g/e$. Now we will use the sum angle formulae for cosine and sine to simplify. Firstly,
\begin{widetext}
\begin{eqnarray}
I(t) = I_0 + \Delta I_0 &\big[& \cos [c_gV_g^{DC}] \cos [c_g(X_1(t)\cos(\Omega_1t) + Y_1(t)\sin(\Omega_1 t) + X_2(t)\cos(\Omega_2 t + \phi) + Y_2(t)\sin(\Omega_2 t + \phi))] \nn \\
&-& \sin [c_gV_g^{DC}] \sin [c_g(X_1(t)\cos(\Omega_1t) + Y_1(t)\sin(\Omega_1 t) + X_2(t)\cos(\Omega_2 t + \phi) + Y_2(t)\sin(\Omega_2 t + \phi))] \big] \nn \\
= I_0 + \Delta I_0 f(t).
\end{eqnarray}
Now, because $V_g^{DC}$ is selected to be at point A, $\cos [c_gV_g^{DC}]=1$ and $\sin [c_gV_g^{DC}]=0$. This simplifies the fluctuation term in the above expression to: 
\begin{eqnarray}
f(t) &=& \cos [c_gX_1(t)\cos(\Omega_1t) + c_gY_1(t)\sin(\Omega_1 t) + c_gX_2(t)\cos(\Omega_2 t + \phi) + c_gY_2(t)\sin(\Omega_2 t + \phi))] \nn \\
&=& \big[ \cos(c_gX_1\cos\Omega_1t)\cos(c_gY_1\sin\Omega_1t) - \sin(c_gX_1\cos\Omega_1t)\sin(c_gY_1\sin\Omega_1t) \big] \cdot \nn \\
&& \big[ \cos(c_gX_2\cos(\Omega_2t + \phi))\cos(c_gY_2\sin(\Omega_2t + \phi)) - \sin(c_gX_2\cos(\Omega_2t + \phi))\sin(c_gY_2\sin(\Omega_2t + \phi)) \big]  \nn \\
&&- \big[ \sin(c_gX_1\cos\Omega_1t)\cos(c_gY_1\sin\Omega_1t) + \cos(c_gX_1\cos\Omega_1t)\sin(c_gY_1\sin\Omega_1t) \big] \cdot \nn \\
&& \big[ \sin(c_gX_2\cos(\Omega_2t + \phi))\cos(c_gY_2\sin(\Omega_2t + \phi)) + \cos(c_gX_2\cos(\Omega_2t + \phi))\sin(c_gY_2\sin(\Omega_2t + \phi)) \big]
\label{eq:f1}
\end{eqnarray}
where we have simply used the sum angle formulae repeatedly, and in the second line made the time dependence of the quadratures implicit. We want to analyze this expression in terms of its harmonics, and to do so we will expand the nested trigonometric functions in a Fourier-Bessel expansion \cite{Abr.Ste-1972} using:
\begin{eqnarray}
\cos(z\cos\theta) = J_0(z) + 2\sum_{k=1}^\infty (-1)^k J_{2k}(z)\cos(2k\theta) \nn \\
\cos(z\sin\theta) = J_0(z) + 2\sum_{k=1}^\infty J_{2k}(z)\cos(2k\theta) \nn \\
\sin(z\cos\theta) = 2\sum_{k=0}^\infty (-1)^k J_{2k+1}(z)\cos((2k+1)\theta) \nn \\
\sin(z\sin\theta) = 2\sum_{k=0}^\infty J_{2k+1}(z)\sin((2k+1)\theta)
\end{eqnarray}
where $J_n$ is the $n^{th}$ order Bessel function. Note that when we apply this expansion to \erf{eq:f1} the arguments to the Bessel function will be the quadratures of the two voltages; therefore we will get powers of $X_i(t)$ and $Y_i(t)$ to all orders. However, in applications of interest we will assume that these quadratures are small (i.e. $C_gV_g \ll e$), and thus ignore contributions from higher powers. Since for small $z$, $J_n(z) \to z^n/2^nn!$ \cite{Abr.Ste-1972}, we can truncate the expansions at $J_1$ if we only consider first order contributions from the quadratures. Doing this yields:
\begin{eqnarray}
f(t) &\approx& J_0(c_gX_1)J_0(c_gY_1)J_0(c_gX_2)J_0(c_gY_2) \nn \\
&&- [2J_1(c_gX_1)J_0(c_gY_1)\cos(\Omega_1t) + 2J_0(c_gX_1)J_1(c_gY_1)\sin(\Omega_1t)]\cdot \nn \\
&& ~~ [2J_1(c_gX_2)J_0(c_gY_2)\cos(\Omega_2t + \phi) + 2J_0(c_gX_2)J_1(c_gY_2)\sin(\Omega_2 t + \phi)]
\end{eqnarray}
Here we have also ignored the terms oscillating at $2\Omega_1$ and $2\Omega_2$ because in applications of interest these high frequency terms will not be detectable. Now expanding $J_0$ and $J_1$ to first order in their arguments,
\begin{eqnarray}
f(t) &\approx& 1 - [2c_gX_1\cos(\Omega_1t) + 2c_gY_1\sin(\Omega_1t)] \cdot [2c_gX_2\cos(\Omega_2t + \phi) + 2c_gY_2\sin(\Omega_2t + \phi)]
\end{eqnarray}
Finally, going back to the expression for the source-drain current, we get
\begin{eqnarray}
I(t) \approx I_0 + \Delta I_0 [ 1 &-& 2c_g^2 X_1X_2 (\cos( (\Omega_1-\Omega_2)t - \phi)+\cos( (\Omega_1+ \Omega_2)t + \phi)) \nn \\
&+& 2c_g^2 X_1Y_2 (\sin( (\Omega_1-\Omega_2)t - \phi)-\sin( (\Omega_1+ \Omega_2)t + \phi)) \nn \\
&-& 2c_g^2 Y_1X_2 (\sin( (\Omega_1-\Omega_2)t - \phi)+\sin( (\Omega_1+ \Omega_2)t + \phi)) \nn \\
&-& 2c_g^2 Y_1Y_2 (\cos( (\Omega_1-\Omega_2)t - \phi)-\cos( (\Omega_1+ \Omega_2)t + \phi))~]
\end{eqnarray}
Now if we ignore the high frequency terms (in applications of interest these frequencies will be above the bandwidth of the intrinsic SET response), we obtain the final expression for the source-drain current:
\begin{eqnarray}
\label{eq:sd_current}
I(t) \approx \tilde{I_0} - 2 \Delta I_0 c_g^2 &[& X_1X_2 \cos( (\Omega_1-\Omega_2)t - \phi) - X_1Y_2 \sin( (\Omega_1-\Omega_2)t - \phi) \nn \\
&&+ Y_1X_2 \sin( (\Omega_1-\Omega_2)t - \phi) + Y_1Y_2 \cos( (\Omega_1-\Omega_2)t - \phi) ~]
\end{eqnarray}
where $\tilde{I_0} \equiv I_0 + \Delta I_0$ is the DC component, and the remaining signal is at the $\Omega_1-\Omega_2$ sideband frequency. This is the expression for the source-drain current at low frequencies and under the assumption that the gate voltage is small. Note that the SET has behaved as a mixer to downconvert the gate voltage signals to the intermediate frequency $|\Omega_1 - \Omega_2|$. We will use this property in the next section when we examine high fidelity measurement strategies for circuit QED.
\end{widetext}

\subsection{SET measurement bandwidths}
As mentioned above, under the appropriate operating conditions the SET has a very high charge sensitivity, and this has led to several applications for it as an electrometer \cite{Kel.Eic.etal-1999, Dre.Ji.etal-1994, Mak.Sch.etal-2001, Goa-2004}. However, the SET has severe limitations when it comes to the measurement of high frequency signals. The intrinsic bandwidth of an SET is high as it is set by the tunneling rate of electrons through the junctions - this is typically higher than 10 GHz \cite{Rei.Bue-2005}. However, the output bandwidth of an SET is constrained to the order of kilohertz because of the high resistance and the parasitic capacitance of the cabling that brings the SET output to room temperature \cite{Pet.Wah.etal-1996}. Typical values for the lead resistance and cable capacitance to the room temperature amplifier are 100k$\Omega$ and 1nF, thus giving an RC time constant of approximately $10^{-4}$, which is much too large to carry radio frequency signals without distortion. This bandwidth limitation of the SET makes it ill suited as a measuring device for circuit QED because of the high frequencies of the signals employed in the architecture: the carrier frequency is the resonance frequency of the microwave cavity, about 10 GHz; and the bandwidth of the signal can be as much as the vacuum Rabi frequency of the qubit, about 100 MHz \cite{Bla.Hua.etal-2004}. Thus the output bandwidth of a typical SET is poorly matched to circuit QED frequencies.

There are two well-known strategies to overcome the bandwidth limitations of an SET. The first is to use the non-linear current-voltage response of the SET to downconvert the signal from a large carrier frequency to an intermediate frequency by mixing with a local oscillator \cite{Kno.Yun.etal-2002}. This allows measurements of signals at a wide range of carrier (center) frequencies, but the measurement bandwidth is still limited by the limited dynamic range of the output leads. 

The second strategy is to embed the SET in a high frequency resonant tank circuit, to create what has been dubbed the RF-SET, and measure the reflected (or transmitted) power from this tank circuit \cite{Sch.Wah.etal-1998}. In this configuration, the SET acts as a variable resistor within an RLC circuit and the SET gate voltage effects the reflected power from the circuit by modifying the damping parameters of the tank circuit. By designing the tank circuit to impedance match the SET and output lines, the SET measurement bandwidth can be increased by more than two orders of magnitude to over 100 MHz \cite{Sch.Wah.etal-1998}. Due to the mixing properties of the SET, the spectrum of the reflected power from the tank circuit has components at the intermediate frequencies $|f_g \pm f_{LC}|$ where $f_g$ is the frequency of the gate voltage and $f_{LC}$ is the resonance frequency of the tank circuit. Thus the RF-SET configuration has the dual effect of increasing the output bandwidth, and downconverting a high carrier frequency signal to an intermediate carrier frequency. But note that the downconversion in this case is to a fixed intermediate frequency (set by the tank circuit parameters) unlike in the previous case where the intermediate frequency could be changed by the local oscillator frequency choice.

Recently, Swenson \textit{et. al} have combined both of these innovations, the local oscillator mixing and the RF-SET, to create an electrometer that has a large output bandwidth as well as a large, tunable center frequency \cite{Swe.Sch.etal-2005}. In the next section we will describe two ways in which the electrometer of Swenson \textit{et. al} can be used to perform measurements in circuit QED. 

\section{Two proposals for high fidelity measurement}
\label{sec:proposals}
In this section we propose two strategies for performing high fidelity QND measurements of the qubits in circuit QED. Both schemes involve using an RF-SET as a solid-state mixer that can be fabricated onto the same circuit as the qubits and waveguide. From Sec. \ref{sec:cqed} we see that the microwave field transmitted from the cavity contains information about the state of the qubits because of the intra-cavity coupling. However, using this transmission line directly as a gate to an RF-SET would not give us the information we need because that would simply measure the intensity of the field, and the qubit state is encoded in the amplitude and phase of the field. This is a standard problem in quantum optics (because photodetectors detect intensity) and there the solution is to mix the signal with a local oscillator. The same technique is used in circuit QED \cite{Bla.Hua.etal-2004}, however the mixing is done at room temperature. In the following we describe how to do this mixing on the circuit using an RF-SET.

\subsection{Scheme 1: Using a dual-gate SET}
\label{sec:dualgate_set}
\begin{figure}
\includegraphics[scale=0.7]{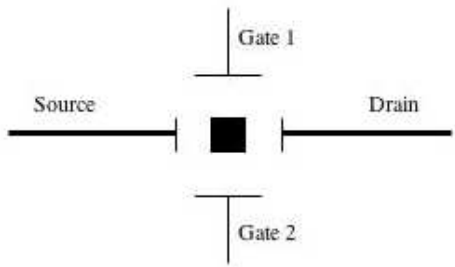}
\caption{Schematic of a dual gate SET. The black box in the middle is the island which is capacitively coupled to the source and drain leads and also the gates. The gate voltage for a dual gate SET is simply the sum of the voltages applied to the two gates.} \label{fig:set}
\end{figure}

We can perform the mixing with a local oscillator using a dual gate SET at the core of the RF-SET. Figure \ref{fig:set} is a schematic of a dual gate single electron transistor. We will assume the source-drain voltage is set such that the SET is in the strong tunneling regime. To perform heterodyne mixing using this device we apply the signal voltage, $V_s$ of frequency $\omega_r$, to one gate and a local oscillator, $V_{LO}$ of frequency $\omega_{LO}$, to the other gate. Then the source-drain current is modulated by both gate voltages. We also bias the SET such that it is operating at point A of Fig. \ref{fig:set_resp}. Then the source-drain current is:
\begin{equation}
I(t) = I_0 + \Delta I_0 \cos \big( \frac{2\pi C_g}{e} (V_g^{DC} + V_s + V_{LO}) \big)
\end{equation}
Then under the assumption that both $V_s$ and $V_{LO}$ are small, we can follow the calculations in Sec. \ref{sec:setmeas} to obtain an expression for this current at the sideband frequency $\omega_r-\omega_{LO}$, in terms of the quadratures of the voltages:
\begin{eqnarray}
I(t) &\approx& \tilde{I_0} - \frac{8\pi^2 \Delta I_0 C_g^2}{e^2} \big[ X_sX_{LO} \cos( (\omega_r-\omega_{LO})t - \phi) \nn \\
&& ~~~~- X_sY_{LO} \sin( (\omega_r-\omega_{LO})t - \phi) \nn \\
&& ~~~~+ Y_sX_{LO} \sin( (\omega_r-\omega_{LO})t - \phi) \nn \\
&& ~~~~+ Y_sY_{LO} \cos( (\omega_r-\omega_{LO})t - \phi) ~\big]
\end{eqnarray}
where $X_s, Y_s$ ($X_{LO}, Y_{LO}$) are the quadratures of the signal (local oscillator), and $\phi$ is the phase difference between the two voltages. In addition, because we are using an RF-SET the output signal has components at $|\omega_{LC} \pm (\omega_r-\omega_{LO})|$ where $\omega_{LC}$ is the tank circuit resonant frequency. At this point we want to choose $\omega_{LO} \approx \omega_r$ and thus have an output signal around the $\omega_{LC}$ center frequency. Therefore let $\omega_{LO} = \omega_r$ and consider the phase choice $\phi=0$. Then considering just the change in the current from the static value ($\tilde{I_0}$) we have
\begin{eqnarray}
\Delta I(t) \approx - \frac{8\pi^2 \Delta I_0 C_g^2}{e^2} &\big[& X_sX_{LO} + Y_sY_{LO} ~\big]
\end{eqnarray}
Expanding the quadrature operators in terms of mode annihilation and creation operators ($X_s = a+a\dg, Y_s = i(a-a\dg), X_{LO}=b+b\dg, Y_{LO} = i(b-b\dg$), where $a$ is the cavity output mode annihilation operator whose explicit expression in terms of the input mode was derived in section \ref{sec:cqed} -- i.e. $a_{out}$ in \erf{eq:inp_oup}) and considering the expectation value of this current under an arbitrary state, we have
\begin{eqnarray}
\langle\Delta I(t)\rangle \approx -\frac{16\pi^2 \Delta I_0 C_g^2}{e^2} \langle a\dg b + b\dg a \rangle 
\end{eqnarray}
at $\omega = \omega_{LC}$. Now finally, if we prepare the local oscillator in a coherent state $\ket{\beta}$ (for $\beta$ real because the relative phase difference between the local oscillator and the signal has already been accounted for by the choice of $\phi$) then we have
\begin{eqnarray}
\frac{\langle\Delta I(t)\rangle}{|\beta|} \approx -\frac{16\pi^2 \Delta I_0 C_g^2}{e^2} \langle a + a\dg \rangle 
\end{eqnarray}
That is, the change in the average monitored current is proportional to the $X$ quadrature of the signal. Similarly, if we had chosen the phase to be $\phi=\pi/2$ the average current would have been proportional to the $Y$ quadrature of the signal voltage. 

Using such a mixer with the source signal being the transmitted field from the circuit QED transmission line will allow one to detect the phase shift of the field mode, and thus the state of the qubits. The advantage of such a scheme is that measurement apparatus, in this case the RF-SET and the local oscillator, can be fabricated directly onto the same circuit as the qubits and resonator and operated at the cryogenic level (e.g. \cite{Seg.Leh.etal-2002}). This would lead to a higher fidelity measurement signal.

\subsection{Scheme 2: Engineering a mixer from waveguides}
\label{sec:trans_mixer}
\begin{figure}
\includegraphics[scale=0.7]{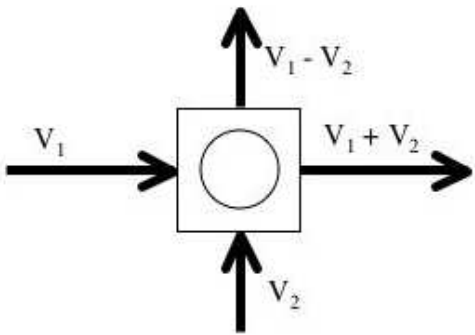}
\caption{Schematic of the transmission line coupling that mimics the behaviour of a beam-spiltter in optics. $V_1$ and $V_2$ are the input signals and the outputs are the sum and difference of these.} \label{fig:resonator_coupling}
\end{figure}

Our second scheme for achieving a high fidelity QND measurement of the qubits involves engineering a coplanar waveguide coupling to mimic a quantum optics beam-splitter. A schematic of such a setup is shown in Figure \ref{fig:resonator_coupling}. The box in the middle generates the sum and difference of the signals provided as input. In the circuit QED model, all the signals are carried on coplanar waveguides and therefore the thick lines represent these waveguides while the arrows indicate the direction of signal propagation. Such a coupling can be engineered through the use of well known four-port hybrid devices, for example, the \textit{rat-race hybrid} \cite{Lee-2004, Poz-2005, Cha-1994}. One of the input waveguides will be the transmission line that couples to the qubits (i.e. the `cavity'), and the other will be a new coplanar waveguide that carries a local oscillator mode. 

Given such a coupling, we can use a single-gate RF-SET to perform intensity measurements of the sum (or difference) signal. That is, the sum (or difference) output signal of the four-port hybrid is made the gate signal of a single gate RF-SET operating in the strong tunneling regime and at point A on Figure \ref{fig:set_resp} (i.e. the DC bias voltage is selected so that the current-voltage curve is at an extremum). This situation then becomes exactly the same as a dual gate SET with the source and local oscillator each being applied to separate gates. In this case $V_1$ takes the place of the signal voltage, $V_2$ takes the place of the local oscillator, and as before the sum of the voltages modulates the source-drain current. Therefore this setup becomes entirely analogous to the RF-SET mixer measurement of section \ref{sec:dualgate_set}, and the analysis from that section can be repeated: if we choose the frequencies of $V_1$ and $V_2$ to be equal, and both voltages to be small, quadrature measurements of the cavity output mode can be made at the tank circuit resonant frequency $\omega_{LC}$. 

The fabrication of SETs with multiple, well characterized gates is a challenging exercise, and therefore this scheme will be easier to implement than the scheme of section \ref{sec:dualgate_set} if the hybrid coupler is easier to engineer than a dual-gate SET. Issues such as cross-coupling, losses and asymmetries at the SET gates are less important with this second scheme. The ideal hybrid coupler is a lossless, unitary device that is likely to be much easier to characterize than a multi-gate SET.

Finally, we note that Mariantoni \textit{et. al.} have recently proposed the use of an on-chip hybrid ring coupler to perform homodyne measurements in a manner very similar to the method above \cite{Mar.Sto.etal-2005}.

\section{A feedback protocol}
\label{sec:fb}
The availability of high fidelity measurement enables the implementation of quantum feedback protocols, and in this section we describe one simple feedback protocol to deterministically generate and stabilize entanglement between two qubits in the circuit-QED architecture. 

As mentioned in section \ref{sec:cqed}, in the dispersive regime the interaction between the microwave cavity mode and a qubit takes the form $H_{int} = (\hbar g^2/\Delta) \sigma_z a\dg a$. If we have two qubits both coupled to the same microwave mode through the Jaynes-Cummings interaction, then the effective Hamiltoninan in the dispersive regime is well approximated by \cite{Ger.Kni-2005, Zhe.Guo-2000}:
\begin{eqnarray}
H_{disp} &\approx& \big( \omega_r + 2\chi J_z \big) a\dg a + \big( \omega_a + \chi \big) J_z \nn \\
&& + \chi \big( \sigma_1^+\sigma_2^- + \sigma_1^-\sigma_2^+ \big)
\end{eqnarray}
where $\hbar$ has been set to 1, $J_z = \frac{1}{2}(\sigma_1^z + \sigma_2^z)$, $\chi \equiv g^2/\Delta$, and $\sigma_i^+$ and $\sigma_i^-$ are the qubit raising and lowering operators for qubit $i$. The last term is an exchange coupling between the qubits that is induced by their interaction with a common mode. It is convenient to work in an interaction picture and simplify the Hamiltonian to:
\begin{eqnarray}
H_{int} &\approx& 2\chi J_z a\dg a + \chi \big( \sigma_1^+\sigma_2^- + \sigma_1^-\sigma_2^+ \big)
\label{eq:tc_disp}
\end{eqnarray} 
The qubit-field interaction can be viewed as one where the field measures a collective property of the qubits. It has been noted by several authors that such an interaction combined with a monitoring of the output field can probabilistically generate an entangled state between the two qubits \cite{Ple.Hue.etal-1999, Zhe.Guo-2000, Nic.Nap.etal-2004}. Additionally, in a series of elegant papers, Mabuchi \textit{et. al.} have shown that it is possible to \textit{deterministically} generate entanglement in a cavity QED system (with a similar Hamiltonian) by employing feedback \cite{Ger.Sto.etal-2004, Sto.Han.etal-2004, Han.Sto.etal-2005}. We will show that it possible to do the same in circuit QED.

The motivation for introducing feedback in order to generate and stabilize entanglement comes from the following simple observation. Consider the evolution of an initial product state of two qubits polarized in the x direction and a coherent state of the field:
\begin{eqnarray}
&& (\ket{00} + \ket{01} + \ket{10} + \ket{11})\otimes \ket{\alpha} \nn \\
&\rightarrow& \ket{00}\ket{\alpha e^{-i2\chi t}} + e^{-i\chi t} \ket{01}\ket{\alpha} + \nn \\
&& ~~~~~~ e^{-i\chi t}\ket{10}\ket{\alpha} + \ket{11}\ket{\alpha e^{i2\chi t}}
\end{eqnarray}
where the arrow indicates unitary evolution for time $t$ under the interaction Hamiltonian \erf{eq:tc_disp}. Notice that the field acquires a phase shift if the qubits are in the $\ket{00}$ or $\ket{11}$ state. Thus, if we apply a $J_x = \frac{1}{2}(\sigma_1^x + \sigma_2^x)$ rotation when we see a phase shift in the field, it is possible that the system could be forced into the entangled state $\ket{\phi^+} = \frac{1}{\sqrt{2}}(\ket{01} + \ket{10})$. We will show that this is indeed the case.

The first step in modeling the dynamical system is to include the driving and damping of the microwave cavity mode. This results in the following master equation for the system:
\begin{eqnarray}
\label{eq:mast_eq_meas}
\frac{d\rho}{dt} = -i[H,\rho] + \kappa \mathcal{D}[a]\rho
\end{eqnarray}
where $\kappa$ is the rate of damping of the cavity, and the Hamiltonian with the added driving term is
\begin{equation}
\label{eq:ent_ham}
H = \epsilon(a + a\dg) + \chi \big( 2J_z a\dg a + \sigma_1^+\sigma_2^- + \sigma_1^-\sigma_2^+ \big)
\end{equation}
Now if we perform a homodyne measurement of the cavity field in order to detect the phase shift, then in analogy with the one qubit case, we gain information about the $J_z$ eigenvalue of the two-qubit state. The measurement current increment is expressed as \cite{Wis.Mil-1993, Wis-1994, Doh.Hab.etal-2000}
\begin{equation}
dI(t) = \kappa\langle a + a\dg \rangle dt + \sqrt{\kappa}dW(t)
\end{equation}
where $dW(t)$ a random variable that represents the noise; it is a Wiener increment: E[$dW(t)$] = 0, and E[$dW(t)dW(s)$] = $\delta(t-s)dt$ \footnote{E[$\cdot$] denotes the classical expectation value.}. If we assume the measurement is unit efficiency, we can represent the evolution of the system as a pure state evolution (also called an unravelling of \erf{eq:mast_eq_meas} into a stochastic Schr\"{o}dinger equation (SSE)) \cite{Wis.Mil-1993, Wis-1994, Doh.Hab.etal-2000}:
\begin{equation}
\label{eq:sse_meas}
d\ket{\psi (t)} = [-iH - \frac{\kappa}{2} a\dg a] \ket{\psi (t)} dt + dI(t) a \ket{\psi (t)}
\end{equation}
This is an unnormalized version of the SSE, and $H$ the Hamiltonian is from \erf{eq:ent_ham}. This is a filtering equation that modifies the state evolution to include the back-action of the measurements while also enforcing consistency with the observed measurement results.

Now we need to add the feedback. The simplest form of feedback is direct Hamiltonian feedback \cite{Wis-1994a}, where the feedback operator is conditioned by a signal that is linearly related to the current (and only the current) measurement value. Such feedback results in Markovian system dynamics and thus an evolution equation that is particularly simple. However, this type of feedback is poorly suited to tasks such as stabilization because of the amount of noise in the instantaneous measurement signal. Therefore, in order to deterministically generate and stabilize entanglement in our system we will use a state-estimate based feedback mechanism that results in non-Markovian, but generally more stable system dynamics. Essentially the task is to estimate the $J_z$ value of the qubit state with high accuracy and base the feedback upon it. In Refs. \cite{Sto.Han.etal-2004, Han.Sto.etal-2005} Mabuchi \textit{et. al.} investigate the optimality and robustness of such an approach in achieving the same goal in a similar system and conclude favorably on it. We will not do the optimality analysis, but will simply illustrate the validity of the feedback protocol by simulation. 

We already have a noisy, continuous measurement of $J_z$ (the homodyne current measures a phase shift, which is directly proportional to the $J_z$ value of the qubits), and we will refine it into a less noisy estimate of $\langle J_z\rangle$ by two steps: (1) low-pass filter the measurement signal over a small time window, and (2) condition the feedback with a power of the filtered measurement signal. The first step forms a crude, but efficient estimate of the observable $\langle J_z \rangle$, and the second reduces the noise in this estimate before conditioning the feedback with it. In the first step the instantaneous measurement is filtered using a simple low-pass filter to produce the smoothed signal
\begin{equation}
R(t) = \frac{1}{\mathcal{N}} \int_{t-T}^t e^{-\gamma(t-t')} dI(t')
\end{equation}
where the factor $\mathcal{N}$ normalises the smoothed signal to an maximum magnitude of unity. We model the feedback by simply adding a conditioned Hamiltonian evolution term to the SSE of \erf{eq:sse_meas} \cite{Wis-1994a, Wis-1994}:
\begin{eqnarray}
\label{eq:sse_fb}
d\ket{\psi (t)} &=& [-iH - \frac{\kappa}{2} a\dg a] \ket{\psi (t)} dt + dI(t) a \ket{\psi (t)} \nn \\
&-& i\lambda R(t)^P J_x \ket{\psi (t)}dt
\end{eqnarray}
where $P$ is the power to which the smoothed signal is raised before conditioning the feedback, and $\lambda$ parametrises the feedback Hamiltonian strength. Such feedback could be performed in the circuit QED architecture by utilizing the electronic control over the Cooper pair box qubits, or by irradiating the cavity with an additional mode in resonance with the qubit transition frequency as done in Ref. \cite{Wal.Sch.etal-2005}.

At this point, we note that there are a number of free parameters in this SSE: $\epsilon$, the field driving strength; $\chi$, the qubit-field coupling strength; $\kappa$, the field damping rate; $\lambda$, the feedback rate; $T$ and $\gamma$, the parameters of the low pass filter; and finally, $P$, the power to which the smoothed current is raised. We will illustrate the effectiveness of the above feedback protocol in generating a maximally entangled two qubit state by numerically solving the (non-linear, non-Markovian) SSE, \erf{eq:sse_fb}, for the following parameter values: $\epsilon=100, \chi=25, \kappa=100, \lambda=100, P=3$. The time step is $dt=0.0001$, the filter values are $T=2000\times dt$ and $\gamma=0.003$, and the total time simulated is 10 time units (the units of time are arbitrary). The choice of these parameters were fairly arbitrary, however, there are some guiding principles for the relationship between the parameters for the scheme to be effective:
\begin{enumerate}
\item We want a heavily damped cavity so that we have as much information about the qubits as possible. We also want to minimize the long-term entanglement between the qubits and the cavity mode because such entanglement would decrease the entanglement between the qubits themselves. Essentially we want the mode to act like an ideal measurement device. These consideration suggest $\epsilon \approx \kappa \gg \chi$.
\item The feedback should be strong enough to compete with the damping effects of the field. Thus $\lambda \geq \kappa$.
\item The filter should be large enough to reduce the noise to acceptable levels but still be responsive to signal changes.
\item We should raise the smoothed signal to a power such that the noise is reduced while the signal is still preserved. The quantity $P=3$ was arrived at by trial. 
\end{enumerate}
Also, we truncate the harmonic oscillator state space to 25 dimensions for the numerics. This truncation is valid because the strong damping keeps the oscillator state close to the vacuum. The stability of the results derived using this truncation were confirmed by comparing to results using a larger state space.

\begin{figure}
\includegraphics[scale=0.6]{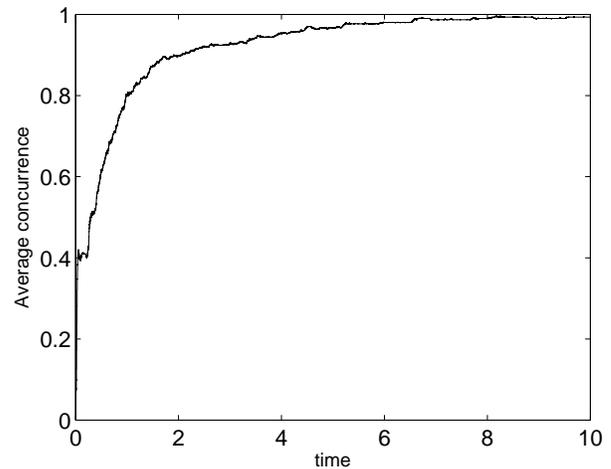}
\caption{The evolution of entanglement averaged over 300 trajectories. The parameters used in the simulation are in the main text. The time units are arbitrary.} \label{fig:ent_evol}
\end{figure}

\begin{figure}
\includegraphics[scale=0.6]{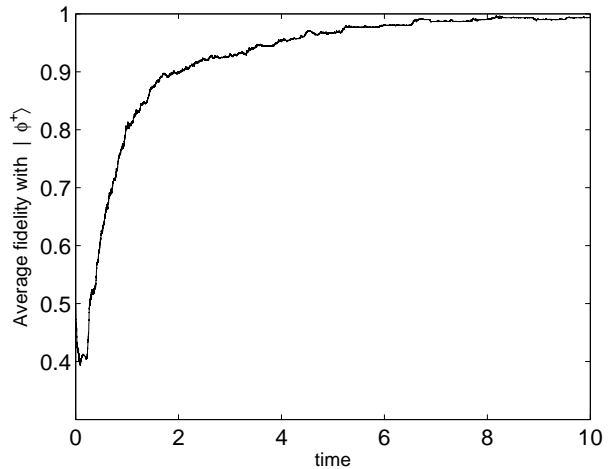}
\caption{The evolution of fidelity with the maximally entangled state $\ket{\phi^+} = \frac{1}{\sqrt{2}}(\ket{01} + \ket{10})$ averaged over 300 trajectories. The parameters used in the simulation are in the main text. The time units are arbitrary.} \label{fig:fid_evol}
\end{figure}

Solving the SSE in \erf{eq:sse_fb} once is not sufficient to make conclusions about the properties of the feedback protocol because that would only determine the performance under one noise record $\{dW(t)\}$. We are more interested in the average properties of the entanglement generation scheme, and thus the SSE is solved 300 times -- 300 trajectories are generated -- all with the initial state $\ket{\psi_0} = (\frac{1}{2}(\ket{00} + \ket{01} + \ket{10} + \ket{11})\otimes \ket{\alpha}$ with $\alpha=3$. Then the average entanglement generated is used as a measure of performance. Figures \ref{fig:ent_evol} and \ref{fig:fid_evol} summarize the results of these simulations. Figure \ref{fig:ent_evol} shows the evolution of the average entanglement (measured using the concurrence measure \cite{Hil.Woo-1997}) of the two qubits, and figure \ref{fig:fid_evol} shows the evolution of the average overlap of the two qubit state with the maximally entangled state $\ket{\phi^+} = \frac{1}{\sqrt{2}}(\ket{01} + \ket{10})$. These figures clearly indicate a convergence to the desired entangled state, in fact we find that 99\% of the 300 trajectories converge onto the $\ket{\phi^+}$ state. Furthermore, as $t\rightarrow \infty$, the convergence percentage approches 100\%. It is important to note that the measurement record indicates the state of the qubits, and therefore we can tell when we have converged on the entangled state.

Thus we have convincing numerical evidence to suggest that the above feedback scheme will converge on the maximally entangled state $\ket{\phi^+}$ in a short time with very high probability.

\section{Conclusion}
\label{sec:conc}
We have presented two schemes for achieving high fidelity single-shot measurements in the circuit QED architecture. As with original proposal \cite{Wal.Sch.etal-2004}, the measurement is done in the Josephson basis and therefore minimizes excursions of the CPB from the optimal charge degeneracy point. Furthermore, the use of dispersive readout and an RF-SET minimizes on-chip power dissipation, an important consideration for solid-state implementations of quantum computing. Both measurement schemes use the RF-SET as a fundamental element, and the possibility of making them quantum-limited hinges crucially on the performance of these SETs. It is conceivable that the ultimate limit to the performance of these measurement schemes is determined by the shot-noise through the SETs. We briefly note that it may be possible to use a Josephson junction as an alternative to the SET to perform the measurements. The sinusoidal current-flux relationship of a large capacitance Josephson junction ($E_J \gg E_c$) \cite{Dev.Wal.etal-2004} could be used to generate the non-linearity needed to mix the signal and local oscillator voltages.

Despite the similarity in the analysis of the two measurement schemes, they are very different physically.  Both bear a resemblance to quantum optics heterodyne measurement, with the second being almost the solid-state counterpart of it. However, a major departure from the quantum optics analogy is the fact that for these SET mixer strategies, we need the signal \textit{and} the local oscillator amplitude to be small. This is a consequence of the ability of SETs to generate non-linearities of all orders. 

High fidelity measurements in the circuit QED architecture will enable many interesting experiments in quantum feedback control, and we have briefly discussed one such experiment to generate and stabilize entanglement between two qubits. The results of Refs. \cite{Ger.Sto.etal-2004, Sto.Han.etal-2004, Han.Sto.etal-2005} suggest that multi-qubit entanglement can also be generated in circuit QED using a similar protocol to the one we have described. In addition to quantum control applications, high fidelity single shot measurements are likely to be crucial for quantum error correction, an essential component of quantum computing.

\section{Acknowledgements}
\label{sec:acks} MS and GJM gratefully acknowledge the support of the Australian Research Council Centre of Excellence in Quantum Computer Technology. HSG is grateful to the Center for Quantum Computer Technology at the University of Queensland for their hospitality during his visit. HSG also 
acknowledges support from the National Science Council, Taiwan under Contract No. NSC 94-2112-M-002-028, and support from the focus group program of the National Center for Theoretical Sciences, Taiwan under contract No. NSC 94-2119-M-002-001.

\bibliography{mybib}

\end{document}